\documentclass[letterpaper]{article} 
\usepackage{aaai25}  
\usepackage{times}  
\usepackage{helvet}  
\usepackage{courier}  
\usepackage[hyphens]{url}  
\usepackage{graphicx} 
\usepackage{natbib}  
\usepackage{caption} 
\usepackage{algorithm}
\usepackage{algorithmic}

\usepackage{newfloat}
\usepackage{listings}
\DeclareCaptionStyle{ruled}{labelfont=normalfont,labelsep=colon,strut=off} 
\floatstyle{ruled}
\newfloat{listing}{tb}{lst}{}
\floatname{listing}{Listing}
\usepackage{subfigure}
\usepackage{xspace}
\usepackage{amsmath}
\usepackage{amssymb}
\usepackage{makecell}
\usepackage{enumitem}
\usepackage{multirow}
\usepackage{booktabs}

\newcommand{\cliper}{{CLIPER}\xspace}

\title{Towards Bridging the Cross-modal Semantic Gap for Multi-modal Recommendation}
\author{
	Xinglong Wu\textsuperscript{\rm 1},
	Anfeng Huang\textsuperscript{\rm 1},
	Hongwei Yang\textsuperscript{\rm 1},
	Hui He\textsuperscript{\rm 1},
	Yu Tai\textsuperscript{\rm 1},
	Weizhe Zhang\textsuperscript{\rm 1}\textsuperscript{\rm 2}
}
\affiliations {
	\textsuperscript{\rm 1}Harbin Institute of Technology\\
	\textsuperscript{\rm 2}Pengcheng Laboratory\\
	xlwu@stu.hit.edu.cn, huanganfeng@stu.hit.edu.cn, yanghongwei@hit.edu.cn,\\
	hehui@hit.edu.cn, taiyu@hit.edu.cn, wzzhang@hit.edu.cn
}

\usepackage{bibentry}

\begin{document}

\maketitle

\begin{abstract}
Multi-modal recommendation greatly enhances the performance of recommender systems by modeling the auxiliary information from multi-modality contents.
Most existing multi-modal recommendation models primarily exploit multimedia information propagation processes to enrich item representations and directly utilize modal-specific embedding vectors independently obtained from upstream pre-trained models.
However, this might be inappropriate since the abundant task-specific semantics remain unexplored, and the cross-modality semantic gap hinders the recommendation performance.

Inspired by the recent progress of the cross-modal alignment model CLIP, in this paper, we propose a novel \textbf{CLIP} \textbf{E}nhanced \textbf{R}ecommender (\textbf{CLIPER}) framework to bridge the semantic gap between modalities and extract fine-grained multi-view semantic information.
Specifically, we introduce a multi-view modality-alignment approach for representation extraction and measure the semantic similarity between modalities.
Furthermore, we integrate the multi-view multimedia representations into downstream recommendation models.
Extensive experiments conducted on three public datasets demonstrate the consistent superiority of our model over state-of-the-art multi-modal recommendation models.
\end{abstract}

%

\section{Introduction}
The utilization of multi-modal side information as a complement to enhance the performance of recommender systems has demonstrated promising foreground and gained widespread adoption.
In order to explore the propagation pattern of multi-modal content consumed by users, Multi-modal Recommender Systems (MMRSs) improve collaborative filtering performance by delving into latent semantic information within multimedia features, with the aim of uncovering content propagation patterns and user interaction preferences based on content consumption records.
Therefore, investigating the relationships between multi-modal content and user preferences remains a fundamental challenge in multi-modal recommendations.

Various MMRS methods have been proposed to bridge the gap between collaborative interaction modeling and multi-media information diffusion pattern analysis.
Early attempts~\cite{he2016vbpr, chen2017acf} aim to incorporate visual information into matrix decomposition in order to establish a connection between interaction records and item contents.
With the prevalence of Graph Neural Networks (GNNs) \cite{wang2019ngcf, he2020lightgcn, wu2023pdagnn}, GNN-based multi-modal recommendation offers a significant advantage in exploring high-order interaction dependencies by integrating multi-modal features with user/item representation (ID embedding), becoming the mainstream solution~\cite{wei2019mmgcn, wei2020grcn, zhang2021lattice, zhou2023freedom}.
To further exploit modality correlations, Self-Supervised Learning is employed as a data augmentation approach to enhance modality-aware recommendation performance \cite{wei2023mmssl, zhou2023bm3}.

Although existing MMRS models have achieved impressive performances by incorporating multi-media features into latent representations of entities (users/items), we contend that these methods treat multimodal information as a fixed input source and primarily focus on propagating the limited-capacity multimodal information.
Consequently, they overlook the vast amount of knowledge contained within the multimodal information itself, leading to limited improvements in performance.
Specifically, most models typically rely on pre-trained visual models (e.g., ResNet \cite{he2016resnet} or ViT \cite{Dosovitskiy2021vit}) and language models (e.g., BERT \cite{Devlin2019bert}) to acquire visual and textual representations respectively.
However, such reliance on accessing multimodal representations may impede recommendation enhancements in the following aspects.
\textit{First, the semantic information of the multi-modality itself is not fully exploited while coarsely absorbing the context as a whole, resulting in a suboptimal utilization of contextual information and thereby constraining the potential for effective multimodal modeling.}
Taking the Amazon-Review\footnote{\url{https://amazon-reviews-2023.github.io/}}~\cite{hou2024blair} dataset as an example, textual descriptions encompass various metadata including \textit{category, brand, price, etc.}
Typical processing overlooks these fine-grained details and directly inputs the data into pre-trained language models, using the generated embeddings as feature vectors for multi-modal recommendation models.
Such modality information modeling neglects fine-grained domain features and may introduce noisy signals into downstream training processes.
\textit{Second, the semantic gap between multiple modalities leads to fragmentation and segregation among uni-modal expressions.}
Specifically, existing multi-modal recommendation methods often handle each uni-modal representation independently without considering the semantic gaps that exist among different modal representations.
Such uni-modality processing without cross-modal joint analysis fails to fully explore the vast amount of available knowledge and thus limits performance improvements.
Despite efforts made by many MMRS models to address representation inconsistencies, we argue that non-unified input of multi-modal representations directly hampers downstream recommendation model performance.

With the recent success of the Contrastive Language-Image Pre-training (CLIP) \cite{radford2021clip} model, a natural solution emerges to address the aforementioned challenges of \textit{exploring deep semantic information} and \textit{bridging the semantic gap between different modalities} based on CLIP.
By leveraging CLIP's remarkable Cross-Modal Alignment (CMA) capability, we delve into the extensive latent semantics within multi-modal user-item interactions and simultaneously align distinct modality representations for better modeling the interplay between modalities.
Specifically, CLIP is a joint vision-language pre-training model that is trained on 400M image-text pairs, endowing it with semantic perception and cross-modality comprehension abilities.
Previous research in various domains (including drawing synthesis \cite{frans2022clipdraw}, image quality assessment \cite{wang2023clipiqa}, image segmentation \cite{luddecke2022clipseg}, etc.) has demonstrated promising performance when utilizing CLIP for cross-modal content processing.
However, surprisingly little effort has been devoted to exploring its potential for cross-modality alignment in multi-modal recommendations.

In this paper, aiming at bridging the cross-modal semantic gap, we propose a \textbf{CLIP} \textbf{E}nhanced \textbf{R}ecommender framework, dubbed \textbf{CLIPER}, for multi-modal recommendation.
In concrete, we initially investigate the fine-grained textual descriptions and segment these textual features in units of \textit{fields}.
By treating each field of textual description as an observation view, we gain a comprehensive understanding of item representations from multiple perspectives.
Subsequently, leveraging the multi-view textual descriptions as prompts, paired with corresponding images as inputs to CLIP, we obtain the pre-trained visual and language representations and multi-view similarity measurement embedding.
Lastly, by integrating different representations through the Fusion Layer, we propose a model-agnostic framework that is compatible with various backbone models for multi-modal recommendation.

Our main contributions are summarized as follows:
\begin{itemize}[leftmargin=*]
	\item We highlight the significance of the cross-modal semantic gap elimination, and utilize the pre-trained inter-modality knowledge to enhance the entity representations.
	To the best of our knowledge, we are the first to introduce multi-view CLIP into multi-modal recommendation scenarios.
	\item We propose \cliper, a model-agnostic framework that leverages multi-view semantic information bridging for representation alignment and enhancement, leading to significant improvements.
	\item Extensive experiments conducted on three real-world datasets using mainstream MMRS backbone models demonstrate the effectiveness and efficiency of our \cliper model.
	The code and datasets used in our work have been made publicly available
	\footnote{\url{https://github.com/WuXinglong-HIT/CLIPER.git}}.
\end{itemize}

\section{Related Work}
In this section, we introduce researches related to our work in two domains, \textit{i.e.}, multi-modal recommendation and cross-modal alignment.

\subsection{Multi-modal Recommendation}
The mainstream MMRS methods can generally be categorized into two groups:
(1) Collaborative Filtering-based methods, and (2) Graph Neural Network-based methods.

\textbf{Collaborative Filtering-based Methods.}
With the success of collaborative filtering \cite{he2017ncf} in Recommender Systems (RSs) \cite{rendle2010fm}, most early multi-modal recommendation models \cite{zhou2023mmrssurvey} dedicate to directly incorporate visual or textural side information into entity ID embeddings to jointly model the modality-enhanced user-item interactions.
Early attempts (e.g., VBPR \cite{he2016vbpr}, DeepStyle \cite{liu2017deepstyle}, ACF \cite{chen2017acf}) utilize deep learning (including Multi-Layer Perceptron, Attention mechanism or Concatenation operation) to integrate the visual or textual context information into the collaborative filtering framework.

\textbf{Graph Neural Network-based Methods.}
Recently, researchers have verified the efficacy of applying Graph Neural Networks in recommendation tasks \cite{wang2019ngcf, he2020lightgcn}.
The primary focus of graph-based models lies in the propagation scheme and data augmentation by constructing auxiliary graphs.
Specifically, some studies (e.g., MMGCN~\cite{wei2019mmgcn}, GRCN~\cite{wei2020grcn}, MGAT~\cite{tao2020mgat}) illustrate the diffusion process of modalities by specifying and refining the modality transmission process.
Subsequently, LATTICE \cite{zhang2021lattice}, FREEDOM \cite{zhou2023freedom}, and DualGNN \cite{wang2021dualgnn} augment data representations by designing item-item or user co-occurrence graphs to further enrich the interplay between modalities.
Some other studies (e.g., BM3 \cite{zhou2023bm3}, MMSSL \cite{wei2023mmssl}) employ a Self-supervised Learning (SSL) paradigm for data augmentation in MMRS, enriching the relationship richness.


\begin{figure*}[!htbp]
	\centering
	\scriptsize
	\subfigure[An illustrative example.]{
		\label{fig:illustration}
		\includegraphics[height=12.5\baselineskip]{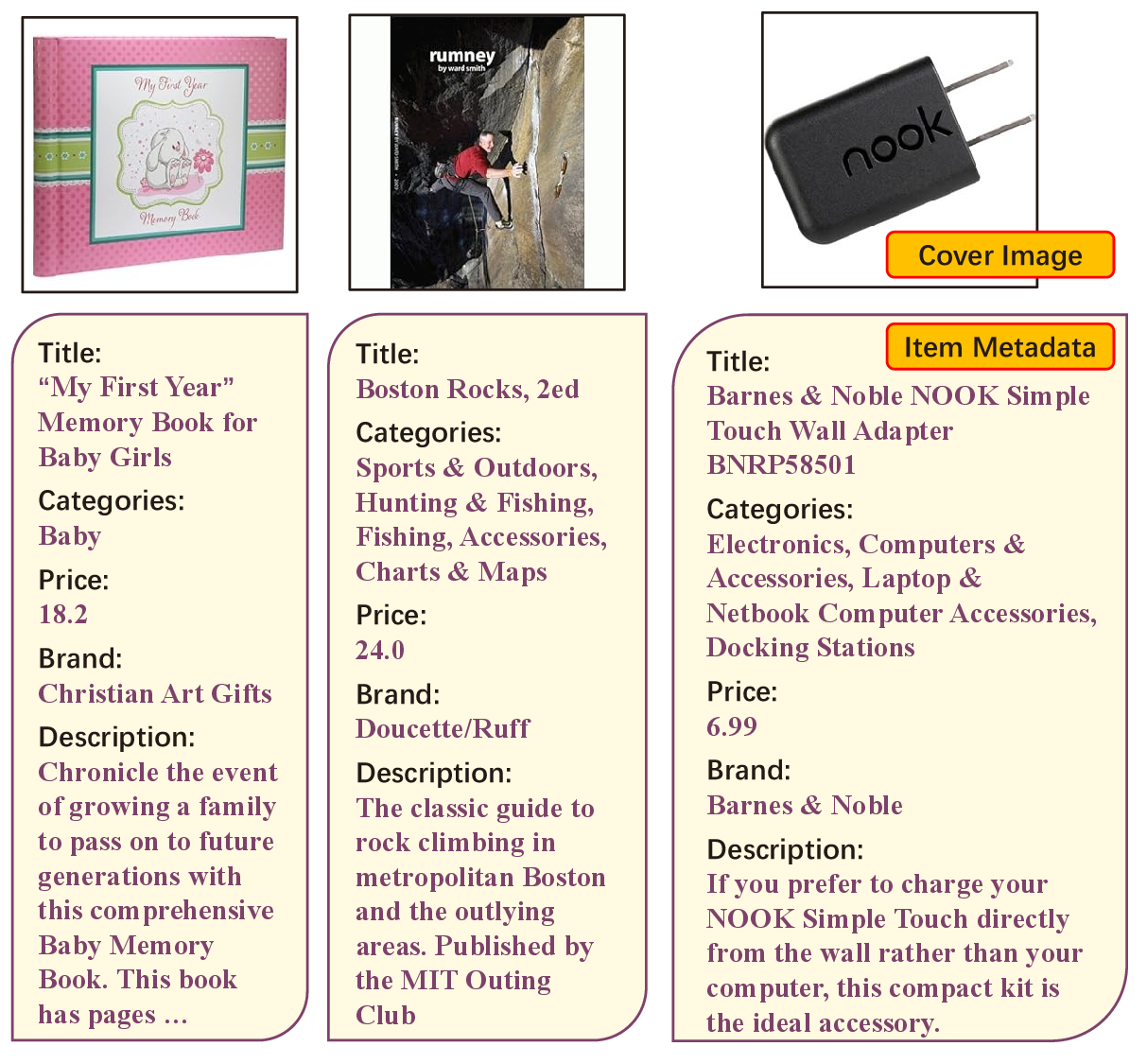}
	}
	\subfigure[Architecture of \cliper.]{
		\label{fig:architecture}
		\includegraphics[height=12.5\baselineskip]{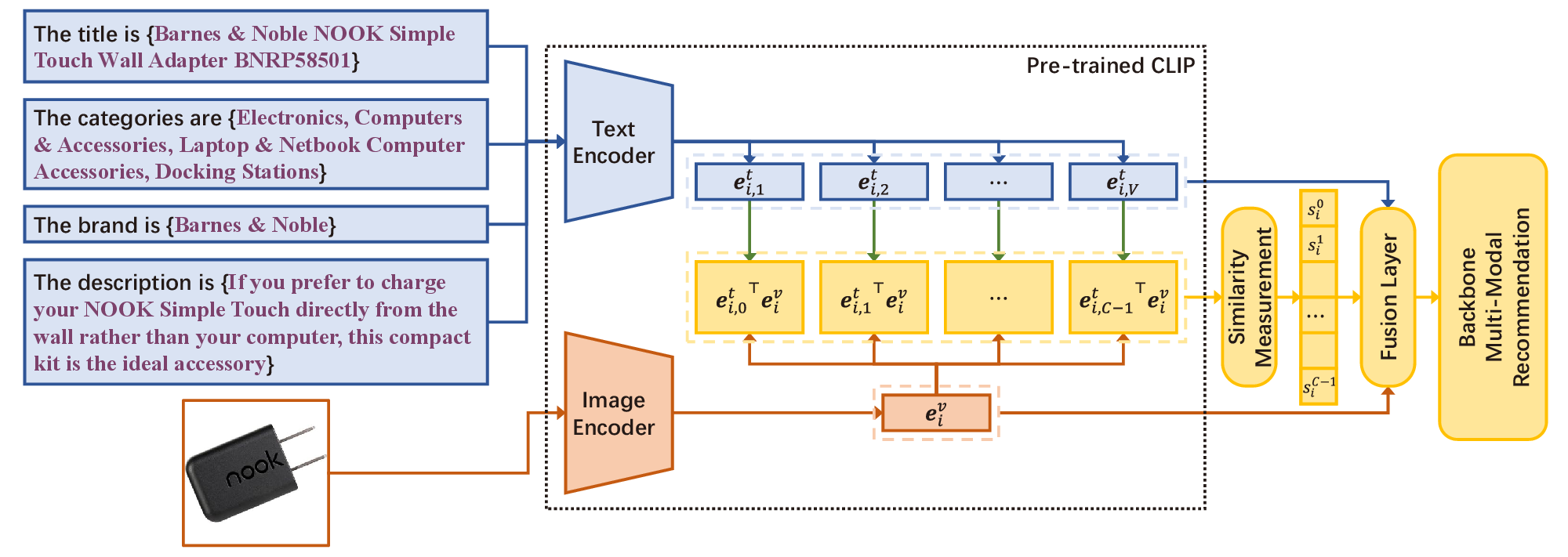}
	}
	\caption{Schematic illustration of the workflow of our proposed \cliper.}
	\label{fig:methodology}
\end{figure*}

\subsection{Cross-modal Alignment}
As mentioned above, our model leverages the cross-modal alignment representation capability to enrich the semantics of uni-modal data and jointly express entities in semantic space from multiple perspectives with the assistance of CLIP \cite{radford2021clip}.
CLIP is a contrastive learning-based neural network that is trained on 400 million image-text pairs, thus possessing the ability of cross-modal semantic extraction.
CLIP has garnered significant attention for various downstream tasks, such as video processing \cite{weng2023openvclip, lei2021clipbert}, semantic segmentation \cite{li2022lseg}, object detection \cite{gu2022vild}, and image synthesis \cite{frans2022clipdraw, vinker2022clipasso}.
However, there is surprisingly little work that harnesses CLIP for multi-modal recommendations and exploits its cross-modality alignment capabilities.

\section{Methodology}
We design our {\textbf{CLIP}} {\textbf{E}}nhanced {\textbf{R}}ecommender (\textbf{CLIPER}) framework for multi-modal recommendation, aiming to bridge the semantic gap between modalities and seamlessly integrate it into various backbone recommendation models.
The overall workflow of our \cliper is illustrated in Figure~\ref{fig:methodology}.
In this section, we first present the problem statement of \cliper, followed by the detailed design of our model.

\subsection{Preliminaries}
\subsubsection{Problem Statement}
Let
\begin{math}
	\mathbf{R} \in \{0, 1\}^{M \times N}
\end{math}
denote interactions between user set
\begin{math}
	\mathcal{U} = \{u_{1}, u_{2}, \cdots, u_{M}\}
\end{math}
and item set
\begin{math}
	\mathcal{I} = \left\{ i_{1}, i_{2}, \cdots, i_{N} \right\}
\end{math},
with each entry $r_{ui} = 1$ denoting the positive interaction between user $u \in \mathcal{U}$ and item $i \in \mathcal{I}$,
where $M=|\mathcal{U}|$ and $N=|\mathcal{I}|$ denote the number of users and items, respectively.
The objective of classical Collaborative Filtering is to predict the user preference for un-interacted items using the prediction score $\hat{r}_{ui}$.
Analogously, multi-modal recommendation aims to address the same problem but incorporates multimodal information $\mathbf{e}_{i}^{m} \in \mathbb{R}^{d_{m}}$ to enhance item modality representations.
Here, $m \in \mathcal{M}$ represents a specific modality; $d_{m}$ denotes the dimension of the representation for modality $m$, and $\mathcal{M}$ denotes the set of modalities.
In this paper, we mainly consider the textural and visual modalities, \textit{i.e.,} $\mathcal{M} = \{t, v\}$.

To sum up, the goal of multi-modal recommendation is to predict the missing user-item interactions, denoted as $\hat{\mathbf{R}}$, by leveraging multi-modal features.
We formalize the process as
\begin{math}
	\hat{\mathbf{R}} = f(\mathcal{U}, \mathcal{I}, \mathbf{R}, \mathbf{E})
\end{math}.

\subsection{Extending CLIP for Multi-modal Recommendation}
\subsubsection{Semantic View Extraction}
\label{sec:semantic view}
To extract comprehensive semantic information at a fine-grained level, we employ a multi-view prompt for semantic exploration.
The item metadata corpus is segmented into units of \textit{field} and divided into $C$ channels as illustrated in Figure~\ref{fig:illustration}, with each channel representing a unique view.
By pairing each individual prompt with the corresponding image, we obtain textual-visual representation tuples
\begin{math}
	(T_{i}^{j}, V_{i})
\end{math},
where $j \in \left\{0, 1, \cdots, C-1\right\}$;
$T_{i}^{j}$ denotes the $j$th textual prompt of item $i$,
and $V_{i}$ denotes the cover image of the item $i$.
Here, $C$ represents the number of views (or channels), with each single view being extracted by a unique prompt template.
Further details on our prompt design can be found below.

\subsubsection{Multi-modal representation Extraction}

The textual representation $\mathbf{e}_{i}^{t}$ and visual representation $\mathbf{e}_{i}^{v}$ for each item $i$ are obtained using the text encoder $E_{t}\left( \cdot \right)$ and image encoder $E_{v} \left( \cdot \right)$ as follows:
\begin{equation}
	\label{eq:encoder}
	\mathbf{e}_{i, j}^{t} = E_{t}(T_{i}^{j}), \qquad
	\mathbf{e}_{i}^{v} = E_{v}(V_{i}).
\end{equation}
Specifically, the cover image $V_{i}$ and the $j$th textual prompt $T_{i}^{j}$ of item $i$ are encoded through the encoders $E_{t}\left( \cdot \right)$ and $E_{v} \left( \cdot \right)$.
Through the contrastive encoders in CLIP, we obtain aligned latent textual embeddings $\mathbf{e}_{i, j}^{t} \in \mathbb{R}^{D}$ and visual embeddings $\mathbf{e}_{i}^{v} \in \mathbb{R}^{D}$ within a unified latent semantic space, where $D=768$ denotes the size of representations in CLIP.

It is worth noting that, unlike conventional studies that separately encode uni-modal representations, the modality-specific representations encoded by each encoder are contrastively aligned in a shared semantic space.
Moreover, these representations contain a wealth of semantic information due to the inclusion of massive image-text pairs during CLIP's pre-training stage.
Specifically, the visual representations offer a more comprehensive depiction by incorporating unselected raw information and intricate details.
On the other hand, textual representations are more precise and focused but lack sufficient content to describe every aspect of the image.
Therefore, by segmenting the textual descriptions into multiple perspectives, we can concentrate on different semantic aspects.
The integration of various views pertaining to the same item results in a more holistic representation.

By encoding both texts and images into the same latent space via the contrastive paradigm, modality-aware representations acquire the semantic perception capability for downstream recommendation tasks.
This is achieved by aligning the latent embedding distribution in the semantic space through a contrastive module that connects the text and image encoders.
Similarity measurement is employed to assess the consistency of semantic expression:
\begin{equation}
	s_{i}^{j} = \frac{ \exp \left( sim \left( \mathbf{e}_{i, j}^{t}, \mathbf{e}_{i}^{v} \right) / \tau \right) }{ \sum_{j^{\prime=0}}^{C-1} \exp \left( sim \left( \mathbf{e}_{i, j^{\prime}}^{t}, \mathbf{e}_{i}^{v} \right) / \tau \right)  },
	\label{eq:similarity}
\end{equation}
where the discriminator function $sim\left( \cdot, \cdot\right): \mathbb{R}^{D} \times \mathbb{R}^{D} \rightarrow \mathbb{R}$ is implemented with the cosine similarity function, i.e.,
\begin{math}
	sim \left( \mathbf{e}_{i, j}^{t}, \mathbf{e}_{i}^{v} \right) = \frac{{\mathbf{e}_{i, j}^{t}}^{\top} \mathbf{e}_{i}^{v}}{\|\mathbf{e}_{i, j}^{t}\| \cdot \|\mathbf{e}_{i}^{v}\|}
\end{math};
$\tau$ is the temperature coefficient to scale the contrast between different modalities.
We integrate all the similarity scores between all the view-specific prompts and the image of the item $i$ and obtain the similarity embedding $\mathbf{s}_{i} = \{{s}_{i}^{j} |_{j=0}^{C-1}\} \in \mathbb{R}^{C}$.

\subsection{Fusion Layer}
The Fusion Layer integrates above encoded diverse representations to form a comprehensive representation for downstream recommendation tasks, encompassing textual representations $\{ \mathbf{e}_{i, j}^{t} \}_{j=0}^{C-1}$, visual representation $\mathbf{e}_{i}^{v}$, and semantic similarity representation $\mathbf{s}_{i}$.

We directly utilize the visual representation $\mathbf{e}_{i}^{v}$ in the downstream recommendation backbone to leverage its rich semantic information derived from CLIP.
However, when it comes to integrating textual representations, text descriptions tend to be more specific and may not fully capture all semantics.
Therefore, we propose four effective integration methods for the fusion layer to handle textual representations: (1) SUM pooling operation, (2) Embedding Concatenation (Concat) operation, (3) Multi-Layer Perceptron (MLP) operation and (4) Self-Attention (SA) operation.

Specifically, for SUM, Concat, and MLP operation, since the latent embeddings share the same semantic space and possess the same vector dimension, we discard the similarity representation for maximum preservation of the original semantics, \textit{i.e.,}
\begin{math}
	SUM: \mathbf{e}_{i}^{t} \leftarrow \sum_{j=0}^{C-1}\mathbf{e}_{i, j}^{t}
\end{math};
\begin{math}
	MLP: \mathbf{e}_{i}^{t} \leftarrow MLP( \| \{\mathbf{e}_{i, j}^{t}\}_{j=0}^{C-1})
\end{math},
where $\|$ denotes concatenation operator, and the output dimension is $D$; we employ a one-layer MLP implementation and utilize $LeakyReLU(\cdot)$ as the activation function.
The detailed implementations for Concatenation and Self-Attention operations are provided below:
\begin{equation}
	\begin{aligned}
		Concat: \mathbf{e}_{i}^{t} &\leftarrow \mathbf{e}_{i, 0}^{t} \| \mathbf{e}_{i, 2}^{t} \| \cdots \| \mathbf{e}_{i, C-1}^{t}, \\
		SA: \mathbf{e}_{i}^{t} &\leftarrow \sum\nolimits_{j=0}^{C-1} \mathbf{e}_{i, j}^{t} \cdot s_{i}^{j}.
	\end{aligned}
	\label{eq:fusion}
\end{equation}

\subsection{Downstream recommendation}
As introduced above, taking the refined latent multi-modal embedding $\mathbf{e}_{i}$ as the semantic information and dumping such pre-trained embeddings as input to the downstream recommendation backbone, we can semantically represent item multi-modality attributes.
The refined latent multi-modal embeddings $\mathbf{e}_{i}^{t}$ and $\mathbf{e}_{i}^{v}$ are utilized as the semantic information, which is then fed into the downstream recommendation backbone.
The extracted representation vector from our \cliper model is seamlessly integrated into subsequent downstream tasks, enabling a plug-and-play approach.
Our model-agnostic design ensures compatibility with any downstream multi-modal recommendation model.

\section{Experiments}
We conduct extensive experiments on three datasets to answer the following four research questions:

\begin{itemize}[leftmargin=*]
	\item \textbf{RQ1.} How do the MMRS models enhanced by \cliper perform compared with the vanilla version?
	\item \textbf{RQ2.} How do the critical components of \cliper impact the recommendation performance?
	\item \textbf{RQ3.} How do the key parameters affect the performance?
	\item \textbf{RQ4.} How does our \cliper model explore the semantic information in multi-modal recommendation scenarios?
\end{itemize}

We first briefly present the experimental settings of our model, followed by the answers to the above questions.

\subsection{Experimental Setups}
\subsubsection{Prompt Design}
\label{sec:prompt}

We segment the attribute corpus into units of ``\textit{field}'' to obtain the fine-grained descriptions of the item.
Subsequently, unnecessary or noisy fields are filtered out, and only rich semantic fields are retained for further processing.
By treating each field as a singular \textit{view}, we fill the value of each view to the corresponding prompt template.
Each field is treated as an individual ``\textit{view}'', and its corresponding prompt template is filled with the respective value.
Some specific prompt templates are provided below:
\begin{itemize}[leftmargin=*]
	\item The product brand is \underline{\ \ \ \ \ \ \ \ \ \ }.
	\item The product categories are \underline{\ \ \ \ \ \ \ \ \ \ }.
	\item The product title is \underline{\ \ \ \ \ \ \ \ \ \ }.
	\item The product description is \underline{\ \ \ \ \ \ \ \ \ \ }.
\end{itemize}

It is worth noting that despite the indispensability of fields such as \textit{title} and \textit{brand} in real-world E-Commerce websites, certain recommendation datasets exhibit low quality by indiscriminately mixing all fields together.
In such cases, apart from laborious data pre-processing, we propose the use of a \textit{global-view} prompt, namely ``The product descriptions are \underline{\ \ \ \ \ \ \ \ \ \ }''.
We additionally include the \textit{global} view in the subsequent experiments by concatenating all the aforementioned individual views.

Moreover, due to the limitation on character length, we truncate the prompt to a maximum of $77$ tokens.
Furthermore, recent studies \cite{radford2021clip, zhou2022coop} have verified that the performance of CLIP could be influenced by the choice of prompts.
We believe that the incorporation of well-designed prompt templates would greatly enhance performance, and further exploration in this area is left for future research.

\begin{table}[t]
	\caption{Statistics about Datasets.}
	\begin{tabular}{ccrcc}
		\toprule
		Dataset  & \# Users & \# Items & \# Interactions & Density \\
		\midrule
		Baby     & 19,445   & 7,050    & 160,792         & 0.117\% \\
		Sports   & 35,598   & 18,357   & 296,337         & 0.045\% \\
		Clothing & 39,387   & 23,033   & 278,677         & 0.031\% \\
		\bottomrule
	\end{tabular}
	\label{table:dataset}
\end{table}

\subsubsection{Datasets}
Three real-world datasets are adopted to verify the effectiveness of our proposed model, namely Amazon-Baby, Amazon-Sports, and Amazon-Clothing \cite{he2016amazondataset, mcauley2015amazondataset}.
The data pre-processing and splitting procedure are meticulously applied following the backbone models \cite{zhou2023freedom, wei2019mmgcn}.
Additionally, we crawl the raw image of each item based on the provided image URL in the item metadata.
We exhibit the dataset statistics in Table \ref{table:dataset}.

\subsubsection{Evaluation Metrics}
We treat the observed interactions as positive targets and regard the remaining ones as negative targets.
After sorting the predicted interaction scores between users and all candidate items in descending order and masking the observed positive ones from the training set, we evaluate the recommendation performance using widely-used metrics $Recall@K$ ($R@K$) and $NDCG@K$ ($N@K$) \cite{he2020lightgcn, wei2024llmrec}, where $K=10, 20, 50$.

\subsubsection{Baseline Models}
To evaluate the effectiveness, we compare the performance with the following baseline models:
\begin{itemize}[leftmargin=*]
	\item MMGCN \cite{wei2019mmgcn}: MMGCN is a classical graph-based MMRS method, which introduces modality-specific graphs for modal-aware propagation and aggregation.
	\item DualGNN \cite{wang2021dualgnn}: DualGNN introduces a user co-occurrence graph to complement the user-item graph, enabling the capture of correlations between users and facilitating the modeling of modality fusion patterns.
	\item LATTICE \cite{zhang2021lattice}: LATTICE, widely recognized as a state-of-the-art MMRS, constructs a semantic item-item graph and effectively propagates modality representations on both user-item and item-item graphs.
	\item SLMRec \cite{tao2023slmrec}: SLMRec proposes a novel SSL-based data augmentation method, facilitating contrastive representation modeling.
	\item FREEDOM \cite{zhou2023freedom}: FREEDOM proposes to freeze the item-item semantic graph and denoise the user-item interaction graph on the basis of LATTICE.
\end{itemize}

\begin{table*}[t]
	\centering
	\caption{Overall Performance Comparisons.}
		\resizebox{\textwidth}{!}{
		\begin{tabular}{ccc|cc|cc|cc|cc|cc}
			\toprule
			\multirow{2}{*}{Dataset}  & \multirow{2}{*}{Metric} & \multirow{2}{*}{K}  & \multicolumn{2}{c|}{MMGCN}       & \multicolumn{2}{c|}{DualGNN}      & \multicolumn{2}{c|}{LATTICE}     & \multicolumn{2}{c|}{SLMRec}       & \multicolumn{2}{c}{FREEDOM}      \\
									  &                         &                     & vanilla                & CLIPER  & vanilla                & CLIPER   & vanilla                & CLIPER  & vanilla                & CLIPER   & vanilla                & CLIPER  \\
			\midrule
			\multirow{12}{*}{Baby}    & \multirow{6}{*}{R@K}    & \multirow{2}{*}{10} & \multirow{2}{*}{3.78}  & \textbf{4.13}    & \multirow{2}{*}{4.48}  & \textbf{5.36}     & \multirow{2}{*}{5.47}  & \textbf{6.29}    & \multirow{2}{*}{5.29}  & \textbf{5.42}     & \multirow{2}{*}{6.27}  & \textbf{6.58}    \\
			&                         &                     &                        & ($\uparrow 9.26\%$)  &                        & ($\uparrow 19.64\%$)  &                        & ($\uparrow 14.99\%$) &                        & ($\uparrow 2.46\%$)   &                        & ($\uparrow 4.94\%$)  \\
			&                         & \multirow{2}{*}{20} & \multirow{2}{*}{6.15}  & \textbf{6.49}    & \multirow{2}{*}{7.16}  & \textbf{8.29}     & \multirow{2}{*}{8.50}   & \textbf{9.39}    & \multirow{2}{*}{7.75}  & \textbf{8.08}     & \multirow{2}{*}{9.92}  & \textbf{10.26}   \\
			&                         &                     &                        & ($\uparrow 5.53\%$)  &                        & ($\uparrow 15.78\%$)  &                        & ($\uparrow 10.47\%$) &                        & ($\uparrow 4.26\%$)   &                        & ($\uparrow 3.43\%$)  \\
			&                         & \multirow{2}{*}{50} & \multirow{2}{*}{11.00} & \textbf{11.69}   & \multirow{2}{*}{12.88} & \textbf{14.60}     & \multirow{2}{*}{14.77} & \textbf{15.66}   & \multirow{2}{*}{12.52} & \textbf{13.03}    & \multirow{2}{*}{16.55} & \textbf{17.47}   \\
			&                         &                     &                        & ($\uparrow 6.27\%$)  &                        & ($\uparrow 13.35\%$)  &                        & ($\uparrow 6.03\%$)  &                        & ($\uparrow 4.07\%$)   &                        & ($\uparrow 5.56\%$)  \\
			& \multirow{6}{*}{N@K}    & \multirow{2}{*}{10} & \multirow{2}{*}{2.00}  & \textbf{2.17}    & \multirow{2}{*}{2.40}   & \textbf{2.94}     & \multirow{2}{*}{2.92}  & \textbf{3.52}    & \multirow{2}{*}{2.90}   & \textbf{2.95}     & \multirow{2}{*}{3.30}   & \textbf{3.54}    \\
			&                         &                     &                        & ($\uparrow 8.50\%$)  &                        & ($\uparrow 22.50\%$)  &                        & ($\uparrow 20.55\%$) &                        & ($\uparrow 1.72\%$)   &                        & ($\uparrow 7.27\%$)  \\
			&                         & \multirow{2}{*}{20} & \multirow{2}{*}{2.61}  & \textbf{2.78}    & \multirow{2}{*}{3.09}  & \textbf{3.70}     & \multirow{2}{*}{3.70}   & \textbf{4.31}    & \multirow{2}{*}{3.53}  & \textbf{3.64}     & \multirow{2}{*}{4.24}  & \textbf{4.48}    \\
			&                         &                     &                        & ($\uparrow 6.51\%$)  &                        & ($\uparrow 19.74\%$)  &                        & ($\uparrow 16.49\%$) &                        & ($\uparrow 3.12\%$)   &                        & ($\uparrow 5.66\%$)  \\
			&                         & \multirow{2}{*}{50} & \multirow{2}{*}{3.59}  & \textbf{3.83}    & \multirow{2}{*}{4.24}  & \textbf{4.97}     & \multirow{2}{*}{4.97}  & \textbf{5.59}    & \multirow{2}{*}{4.50}   & \textbf{4.64}     & \multirow{2}{*}{5.58}  & \textbf{5.91}    \\
			&                         &                     &                        & ($\uparrow 6.69\%$)  &                        & ($\uparrow 17.22\%$)  &                        & ($\uparrow 12.47\%$) &                        & ($\uparrow 3.11\%$)   &                        & ($\uparrow 5.91\%$)  \\
			\midrule
			\multirow{12}{*}{Sports}   & \multirow{6}{*}{R@K}    & \multirow{2}{*}{10} & \multirow{2}{*}{3.70}   & \textbf{3.93}    & \multirow{2}{*}{5.68}  & \textbf{6.39}     & \multirow{2}{*}{6.20}   & \textbf{7.14}    & \multirow{2}{*}{6.63}  & \textbf{6.97}     & \multirow{2}{*}{7.17}  & \textbf{7.73}    \\
			&                         &                     &                        & ($\uparrow 6.22\%$)  &                        & ($\uparrow 12.50\%$)  &                        & ($\uparrow 15.16\%$) &                        & ($\uparrow 5.13\%$)   &                        & ($\uparrow 7.81\%$)  \\
			&                         & \multirow{2}{*}{20} & \multirow{2}{*}{6.05}  & \textbf{6.23}    & \multirow{2}{*}{8.59}  & \textbf{9.43}     & \multirow{2}{*}{9.53}  & \textbf{10.56}   & \multirow{2}{*}{9.90}   & \textbf{10.23}    & \multirow{2}{*}{10.89} & \textbf{11.61}   \\
			&                         &                     &                        & ($\uparrow 2.98\%$)  &                        & ($\uparrow 9.78\%$)   &                        & ($\uparrow 10.81\%$) &                        & ($\uparrow 3.33\%$)   &                        & ($\uparrow 6.61\%$)  \\
			&                         & \multirow{2}{*}{50} & \multirow{2}{*}{10.78} & \textbf{10.84}   & \multirow{2}{*}{13.92} & \textbf{15.24}    & \multirow{2}{*}{15.61} & \textbf{16.68}   & \multirow{2}{*}{15.43} & \textbf{16.08}    & \multirow{2}{*}{17.68} & \textbf{18.75}   \\
			&                         &                     &                        & ($\uparrow 0.56\%$)  &                        & ($\uparrow 9.48\%$)   &                        & ($\uparrow 6.85\%$)  &                        & ($\uparrow 4.21\%$)   &                        & ($\uparrow 6.05\%$)  \\
			& \multirow{6}{*}{N@K}    & \multirow{2}{*}{10} & \multirow{2}{*}{1.93}  & \textbf{2.06}    & \multirow{2}{*}{3.10}   & \textbf{3.43}     & \multirow{2}{*}{3.35}  & \textbf{3.92}    & \multirow{2}{*}{3.65}  & \textbf{3.90}     & \multirow{2}{*}{3.85}  & \textbf{4.16}    \\
			&                         &                     &                        & ($\uparrow 6.74\%$)  &                        & ($\uparrow 10.65\%$) &                        & ($\uparrow 17.01\%$) &                        & ($\uparrow 6.85\%$)   &                        & ($\uparrow 8.05\%$)  \\
			&                         & \multirow{2}{*}{20} & \multirow{2}{*}{2.54}  & \textbf{2.65}    & \multirow{2}{*}{3.85}  & \textbf{4.22}    & \multirow{2}{*}{4.24}  & \textbf{4.81}    & \multirow{2}{*}{4.50}   & \textbf{4.75}     & \multirow{2}{*}{4.81}  & \textbf{5.15}    \\
			&                         &                     &                        & ($\uparrow 4.33\%$)  &                        & ($\uparrow 9.61\%$)   &                        & ($\uparrow 13.44\%$) &                        & ($\uparrow 5.56\%$)   &                        & ($\uparrow 7.07\%$)  \\
			&                         & \multirow{2}{*}{50} & \multirow{2}{*}{3.50}  & \textbf{3.59}    & \multirow{2}{*}{4.93}  & \textbf{5.39}     & \multirow{2}{*}{5.44}  & \textbf{6.04}    & \multirow{2}{*}{5.62}  & \textbf{5.94}     & \multirow{2}{*}{6.18}  & \textbf{6.60}    \\
			&                         &                     &                        & ($\uparrow 2.57\%$)  &                        & ($\uparrow 9.33\%$)   &                        & ($\uparrow 11.03\%$) &                        & ($\uparrow 5.69\%$)   &                        & ($\uparrow 6.80\%$)  \\
			\midrule
			\multirow{12}{*}{Clothing} & \multirow{6}{*}{R@K}    & \multirow{2}{*}{10} & \multirow{2}{*}{2.20}   & \textbf{2.33}    & \multirow{2}{*}{4.68}  & \textbf{5.05}     & \multirow{2}{*}{5.24}  & \textbf{5.41}    & \multirow{2}{*}{4.46}  & \textbf{4.82}     & \multirow{2}{*}{6.29}  & \textbf{6.89}    \\
			&                         &                     &                        & ($\uparrow 5.91\%$)  &                        & ($\uparrow 7.91\%$)   &                        & ($\uparrow 3.24\%$)  &                        & ($\uparrow 8.07\%$)   &                        & ($\uparrow 9.54\%$)  \\
			&                         & \multirow{2}{*}{20} & \multirow{2}{*}{3.52}  & \textbf{3.81}     & \multirow{2}{*}{7.12}  & \textbf{7.65}     & \multirow{2}{*}{7.66}  & \textbf{8.10}     & \multirow{2}{*}{6.87}  & \textbf{7.13}     & \multirow{2}{*}{9.41}  & \textbf{10.37}   \\
			&                         &                     &                        & ($\uparrow 8.24\%$)  &                        & ($\uparrow 7.44\%$)   &                        & ($\uparrow 5.74\%$)  &                        & ($\uparrow 3.78\%$)   &                        & ($\uparrow 10.20\%$) \\
			&                         & \multirow{2}{*}{50} & \multirow{2}{*}{6.21}  & \textbf{6.72}    & \multirow{2}{*}{11.45} & \textbf{12.36}    & \multirow{2}{*}{11.61} & \textbf{12.14}   & \multirow{2}{*}{11.07} & \textbf{11.19}    & \multirow{2}{*}{12.00}    & \textbf{16.24}   \\
			&                         &                     &                        & ($\uparrow 8.21\%$)  &                        & ($\uparrow 7.95\%$)   &                        & ($\uparrow 4.57\%$)  &                        & ($\uparrow 1.08\%$)   &                        & ($\uparrow 35.33\%$) \\
			& \multirow{6}{*}{N@K}    & \multirow{2}{*}{10} & \multirow{2}{*}{1.16}  & \textbf{1.20}    & \multirow{2}{*}{2.52}  & \textbf{2.73}     & \multirow{2}{*}{2.79}  & \textbf{2.88}    & \multirow{2}{*}{2.37}  & \textbf{2.61}     & \multirow{2}{*}{3.41}  & \textbf{3.69}    \\
			&                         &                     &                        & ($\uparrow 3.45\%$) &                        & ($\uparrow 8.33\%$)   &                        & ($\uparrow 3.23\%$)  &                        & ($\uparrow 10.13\%$)  &                        & ($\uparrow 8.21\%$)  \\
			&                         & \multirow{2}{*}{20} & \multirow{2}{*}{1.50}   & \textbf{1.57}    & \multirow{2}{*}{3.14}  & \textbf{3.39}     & \multirow{2}{*}{3.41}  & \textbf{3.57}    & \multirow{2}{*}{2.99}  & \textbf{3.20}      & \multirow{2}{*}{4.20}   & \textbf{4.58}    \\
			&                         &                     &                        & ($\uparrow 4.67\%$) &                        & ($\uparrow 7.96\%$)   &                        & ($\uparrow 4.69\%$)  &                        & ($\uparrow 7.02\%$)   &                        & ($\uparrow 9.05\%$)  \\
			&                         & \multirow{2}{*}{50} & \multirow{2}{*}{2.03}  & \textbf{2.15}    & \multirow{2}{*}{4.00}  & \textbf{4.34}     & \multirow{2}{*}{4.20}   & \textbf{4.38}    & \multirow{2}{*}{3.82}  & \textbf{4.01}     & \multirow{2}{*}{4.44}  & \textbf{5.75}    \\
			&                         &                     &                        & ($\uparrow 5.91\%$)  &                        & ($\uparrow 8.50\%$)   &                        & ($\uparrow 4.29\%$)  &                        & ($\uparrow 4.97\%$)   &                        & ($\uparrow 29.50\%$) \\
			\bottomrule
		\end{tabular}
		}
	\label{table:overall performance}
\end{table*}

\subsubsection{Implementation Details}
We implement \cliper based on PyTorch.
Due to the maximum length limitation of $77$ tokens in CLIP, we opt for the Long-CLIP \cite{zhang2024longclip} as the default backend model, as it allows a longer input length of up to $248$ tokens.
We additionally compare the performance between CLIP and Long-CLIP in the following experiment.
The image encoder in CLIP offers two versions, namely ResNet \cite{he2016resnet} and ViT \cite{Dosovitskiy2021vit}.
The ViT-L/14 and sentence Transformer \cite{vaswani2017attention} are chosen as the visual and textual encoders for CLIP, respectively.
Additionally, the LongCLIP-L version is utilized for LongCLIP.
More detailed implementation details can be found in Appendix \ref{sec: appendix implementation}.

\subsection{Overall Performance Comparison (\textbf{RQ1})}
We conduct extensive experiments on the Amazon-Baby, Amazon-Sports, and Amazon-Clothing datasets and demonstrate performance comparisons based on different backbone models in Table \ref{table:overall performance}.
To save space, we express the percentages relative to their original values.
The improvement percentage is marked with $\uparrow$.
Our conclusions derived from the comparative analysis are as follows:
\begin{itemize}[leftmargin=*]
	\item \textbf{\cliper consistently enhances the performance of MMRSs with a decent margin regardless of the backbone model.}
	In specific, \cliper achieves an average improvement over backbone models of $8.40\%$ and $10.50\%$ in terms of $Recall@K$ and $NDCG@K$ on Amazon-Baby, respectively; an average improvement of $7.17\%$, $8.32\%$ in terms of $Recall@K$ and $NDCG@K$ on Amazon-Sports, respectively; and an average improvement of $8.48\%$, $7.99\%$ w.r.t. $Recall@K$ and $NDCG@K$ on Clothing for all baseline models.
	The performance improvement achieved by FREEDOM-\cliper is particularly noteworthy, as it surpasses that of the vanilla FREEDOM model with a remarkable $35.33\%$ enhancement in terms of $Recall@50$.
	This empirical evidence strongly validates the effectiveness of our proposed \cliper model.
	\item \textbf{The backbone model is attentively enhanced by \cliper across different datasets, showcasing its ability to unlock the modeling potential of various backbone models.}
	Specifically, \cliper-enhanced DualGNN, LATTICE, and FREEDOM achieve the greatest performance improvement on Baby, Sports, and Clothing datasets respectively.
	On the Baby dataset, DualGNN-\cliper achieves a mean improvement over the vanilla model of $16.26\%$ and $19.82\%$ in terms of $Recall@K$ and $NDCG@K$.
	Similarly, on Sports dataset, LATTICE-\cliper achieves an improvement of $10.94\%$, and $13.83\%$ in terms of $Recall@K$ and $NDCG@K$.
	Furthermore, on Clothing, FREEDOM-\cliper realizes an improvement of $18.36\%$, and $15.59\%$ in terms of $Recall@K$ and $NDCG@K$, respectively.
	The consistent enhancements across different baseline models highlight the generalizability of \cliper for various MMRS approaches.
	\item \textbf{FREEDOM-\cliper realizes the best performance.}
	The state-of-the-art baseline model FREEDOM exhibits the highest performance among all models, with both the vanilla version and \cliper-enhanced version of FREEDOM outperforming other MMRS methods.
	The \cliper framework further expands its superiority over other MMRS methods.
	Notably, on Clothing, \cliper achieves a remarkable improvement of $35.33\%$ and $29.50\%$ in terms of $Recall@50$ and $NDCG@50$, respectively.
	This demonstrates the superior semantic comprehension capability of \cliper for longer target sequences.
\end{itemize}

\begin{figure}[t]
	\centering
	\scriptsize
	\subfigure[Recall@K on Baby]{
		\includegraphics[width=0.45\linewidth]{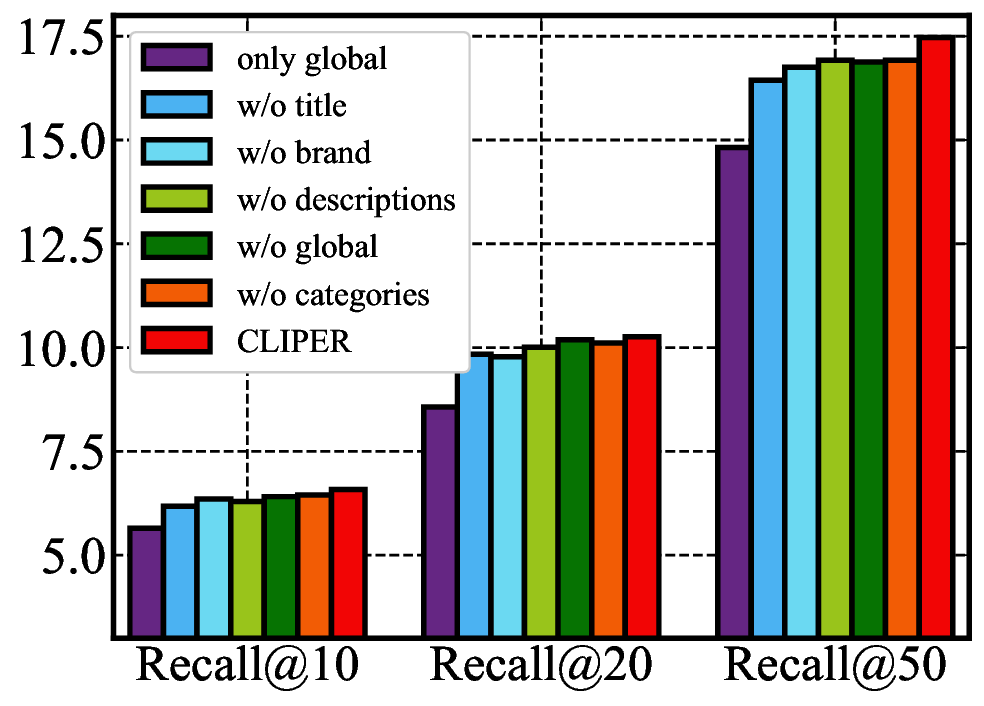}
	}
	\subfigure[NDCG@K on Baby]{
		\includegraphics[width=0.45\linewidth]{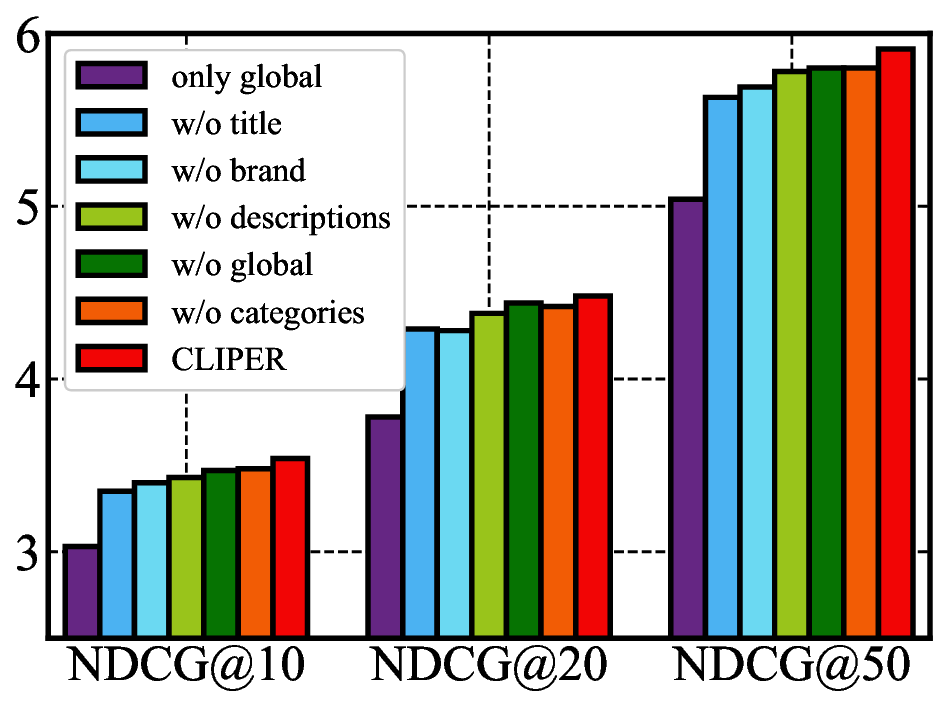}
	}
	\caption{Impact of Individual Views.}
	\label{fig: view}
\end{figure}

\subsection{Model Architecture Study (\textbf{RQ2})}
The impact of various key architecture constitutions, including semantic view ablation and fusion layer implementation, is investigated in this section.
Given that FREEDOM is the most advanced and the highest-performing state-of-the-art model, we employ FREEDOM as our default backbone model for the subsequent study on model architecture and parameter sensitivity analysis without special instructions.

\subsubsection{Impact of individual views}
\label{sec:view impact}
The comprehensive semantic exploration of multi-modal side information is achieved by extracting different semantic views.
To investigate the impact of these views, we conduct ablation experiments on Amazon-Baby by removing specific view $[v]$, denoted as \cliper-$[v]$, including `title', `brand', `categories', `description', and an additional `global' view.
Furthermore, to validate the effectiveness of the multi-view semantic extraction, we explore \cliper with only the `global' view, referred to as \cliper-only-global.
Figure~\ref{fig: view} demonstrates performance comparisons with respect to view ablations.

The consistent outperformance of \cliper over other variants confirms the effectiveness of its fine-grained multi-view semantic exploration. 
Moreover, the varying performance drop indicates a view prioritization as:
title $>$ brand $>$ description $>$ categories.
We argue that this is because `title' contains more refined and condensed information, while `categories' often consist of repetitive and low-value categorical information.
Surprisingly, ablating the `global' view has limited influence, which further validates the effectiveness of our fine-grained multi-view processing.
Furthermore, \cliper-only-global consistently underperforms other variants, demonstrating coarse-grained semantic extraction may impede recommendation performances.

\subsubsection{Impact of Fusion Method}
\label{sec:fusion impact}

\begin{figure*}[t]
	\centering
	\scriptsize
	\begin{minipage}[t]{0.45\textwidth}
		\subfigure[Recall@K on Baby]{
			\includegraphics[width=0.45\linewidth]{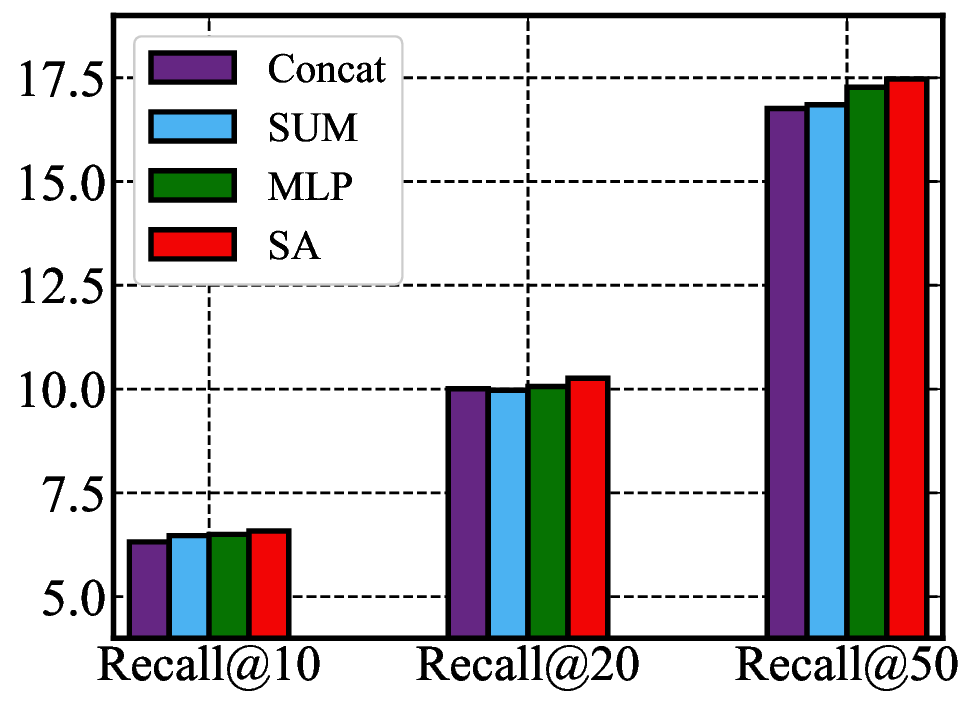}
		}
		\subfigure[NDCG@K on Baby]{
			\includegraphics[width=0.45\linewidth]{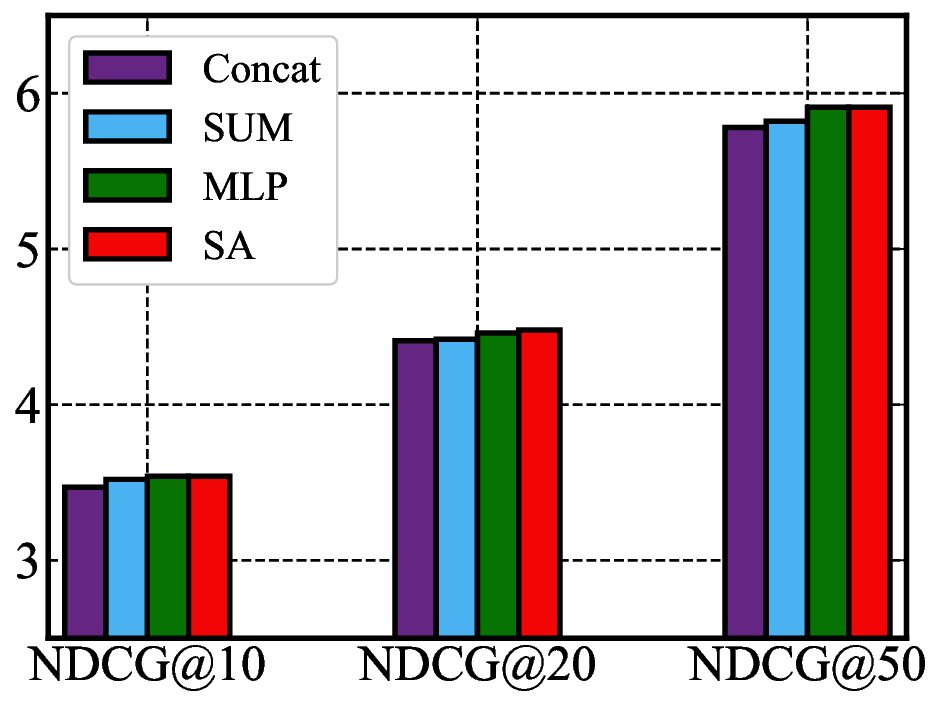}
		}
		\caption{Impact of Fusion Layer.}
		\label{fig:fusion}
	\end{minipage}
	\begin{minipage}[t]{0.45\textwidth}
		\subfigure[Recall@K on Clothing]{
			\includegraphics[width=0.45\linewidth]{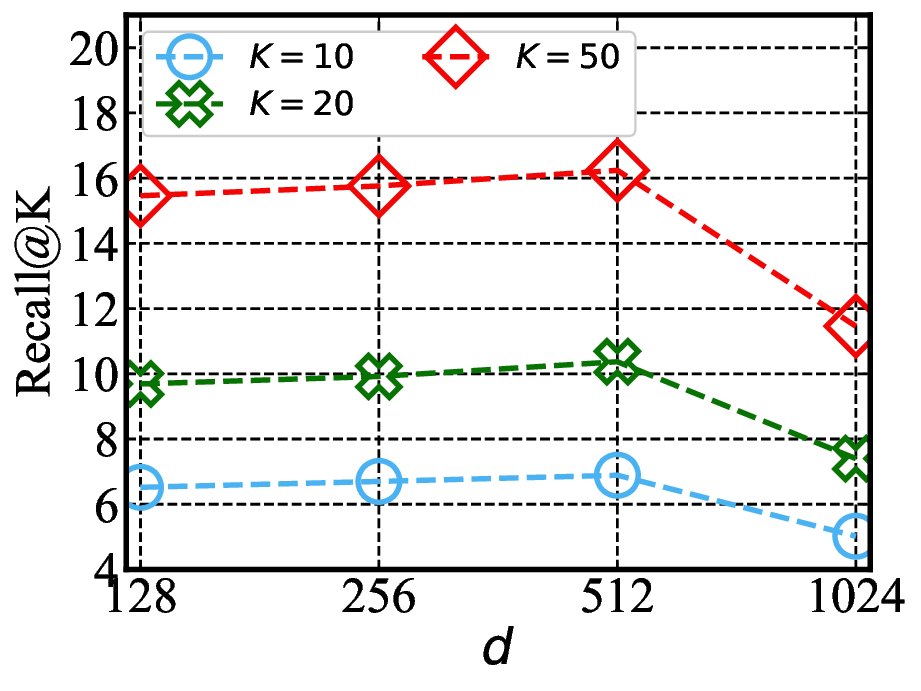}
		}
		\subfigure[NDCG@K on Clothing]{
			\includegraphics[width=0.45\linewidth]{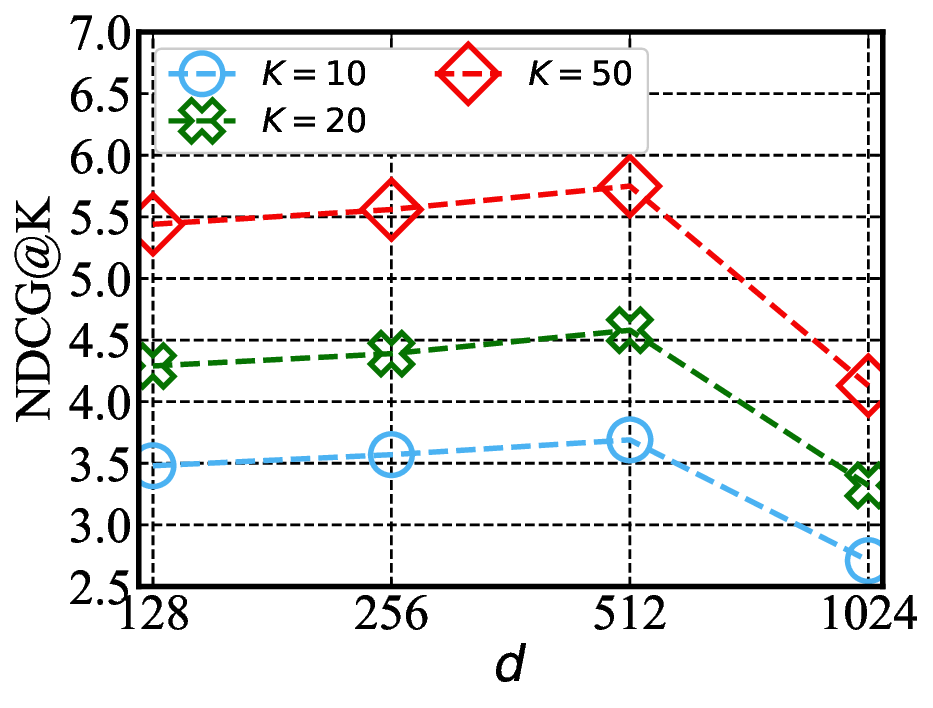}
		}
		\caption{Impact of Embedding Size $d$.}
		\label{fig:dimension}
	\end{minipage}
\end{figure*}

The Fusion Layer integrates various semantic aspects as a cohesive presentation unit for downstream recommendation tasks.
To assess the effectiveness of different fusion implementations, we conduct analysis experiments on the Baby dataset as shown in Figure~\ref{fig:fusion}.
Empirical results from different modality-aware representation fusions demonstrate that the prioritization of fusion implementations is:
SA $>$ MLP $>$ SUM $>$ Concat.
This aligns with our expectations, as the Self-Attention attentively differentiates individual views through similarity scores.
We argue that SUM and Concat cannot rival \cliper-SA due to their coarse-grained integration, which hampers the interplay between uni-view semantic representations and further deteriorates recommendation performance.

\subsection{Parameter Sensitivity Analysis (\textbf{RQ3})}
\label{sec:parameter analysis}

Extensive experiments with respect to key parameters of \cliper, including the embedding size $d$ and temperature coefficient $\tau$, are conducted on the Clothing dataset to optimize the hyper-parameter settings.

We search for the optimal value of $d$ in the range of $\{128, 256, 512, 1024\}$.
As shown in Figure~\ref{fig:dimension}, the performance of \cliper keeps improving as $d$ increases up to $512$, but tends to deteriorate rapidly when $d > 512$.
Setting $d = 512$ yields the best performance.
This observation aligns with our conjecture that increasing the embedding size enhances the capacity of the recommendation model.
Nevertheless, excessively large values may give rise to overfitting issues and hinder the inference ability of the model.

Analysis experiments (c.f. Appendix \ref{sec: appendix temperature tau}) regarding the temperature coefficient $\tau$ demonstrate that setting $\tau=0.1$ yields the best performance in most scenarios.

\subsection{Model Analysis (\textbf{RQ4})}
\label{sec: visualization}

\begin{figure}
	\centering
	\scriptsize
	\includegraphics[width=0.5\linewidth]{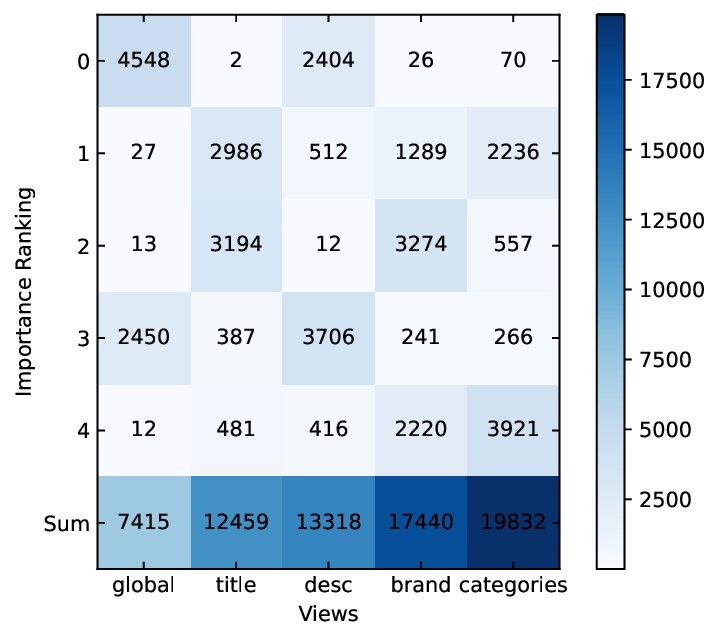}
	\caption{Visualization.}
	\label{fig:heatmap}
\end{figure}

In our work, we propose a multi-view cross-modality alignment approach to bridge fine-grained semantics and leverage similarity measurement as the attention coefficient to integrate view-specific representations.
Since many CLIP-based works utilize similarity vectors for downstream tasks, these vectors inherently possess extracted semantics.
In this section, we attempt to figure out the distribution patterns between individual views and similarity measurement.
Specifically, we calculate similarity scores for each individual view, including the \textit{global} view, for every item in the Amazon-Baby dataset.
We sort these scores in descending order and assign a score $0$ to the most important view and incrementally higher scores to the less important view accordingly.
Additionally, we provide the sum of the product of importance ranking scores and corresponding occurrences in the last row in the heatmap as shown in Figure~\ref{fig:heatmap}.

As is illustrated above, the view priority is as follows: title $>$ brand $>$ description $>$ categories.
We observe that the distribution of importance scores significantly aligns with the priority order.
Specifically, (1) when horizontally comparing each row, the most frequent occurrence of scores ranging from $0$ to $4$ corresponds to \textit{global}, \textit{title}, \textit{brand}, \textit{description} and \textit{categories} respectively, which surprisingly aligns with view priority order. 
The \textit{global} view plays a more important role due to its possession of overall information.
(2) More intuitively, the importance sum reveals an overall distribution where smaller numbers indicate greater effects.
Thus, the overall order of view importance is: \textit{global}, \textit{title}, \textit{description}, \textit{brand} and \textit{categories}, which significantly aligns with empirical observations from the view ablation experiment.
Although no firm conclusions can be drawn solely based on these observations, it can be inferred that there exists a latent semantic priority distributed among multiple modalities.

\section{Conclusion}
In this paper, we propose a novel CLIP-enhanced multi-modal recommendation framework, referred to as \cliper, wherein we establish a bridging approach between cross-modal representations from multiple views and integrate modality-aware representations with semantic similarity coefficients.
This is an initial work to incorporate CLIP into multi-modal recommendations and exploit its ability for cross-modal alignment in order to address the issue of inter-modality semantic gaps.
Specifically, we propose a fine-grained modality-aligned embedding refinement approach through multi-view semantic exploration.
Moreover, the fusion layer attentively integrates each view-specific representation with the similarity measurement coefficient.
Our model-agnostic enhancement framework can seamlessly integrate with existing multi-modal recommendation models in a plug-and-play manner.
Extensive experiments conducted on real-world datasets validate the effectiveness of our proposed model.

\bibliography{ref}

\clearpage
\appendix

\section*{Appendix}

\section{Implementation Details}
\label{sec: appendix implementation}

Regarding the fusion layer design, we explore all four integration methods but primarily employ the SA operation.
The temperature coefficient $\tau$ for similarity measurement is searched within the range of $\{0.05, 0.1, 0.2, 0.5, 0.7\}$.
To ensure fair comparison, the embedding dimensionality $d$ is fixed at a specific value within the range of $\{128, 256, 512, 1024\}$ for downstream recommendation models.

We implement and fine-tune all the baseline models based on their official codes or MMRec\footnote{\url{https://github.com/enoche/MMRec.git}}~\cite{zhou2023mmrssurvey} framework.
The reported performances of all baseline models are well fine-tuned.
We compare the performance of \cliper-enhanced baseline models and their vanilla counterparts.
To ensure accuracy, we select the best-performing results of the vanilla baselines from their reported results in their respective papers, those reported by MMRec, as well as our fine-tuned results.
Specifically, regarding the tuning for each backbone model, we utilize the Adam \cite{kingma2015adam} optimizer.
We employ the early stop strategy for evaluation, where training stops if there is no improvement in the metric $Recall@20$ after $10$ epochs.
The learning rate $lr$ is searched within the range of $\{1, 3, 5\} \times \{e-4, e-5\}$.
By default, the weight decay coefficient $\lambda$ for the $L_{2}$ regularization term is set to $1e-4$.

\section{Detailed Model Architecture and Parameter Study}
\subsubsection{Impact of CLIP}
\label{sec: appendix clip impact}

\begin{figure}[t]
	\centering
	\scriptsize
	\subfigure[Recall@K on Baby]{
		\includegraphics[width=0.48\linewidth]{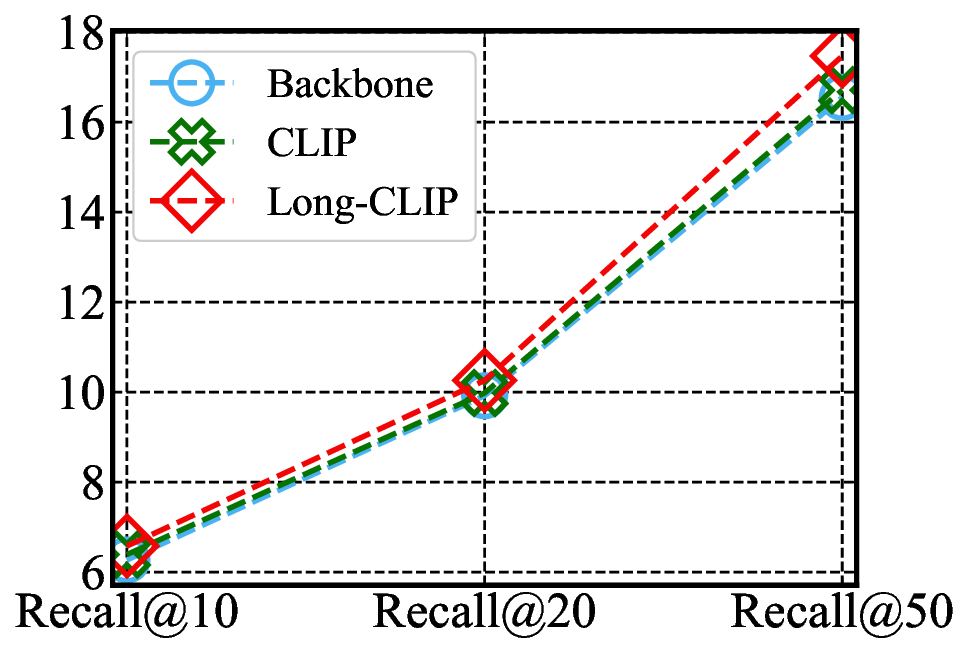}
	}
	\subfigure[NDCG@K on Baby]{
		\includegraphics[width=0.48\linewidth]{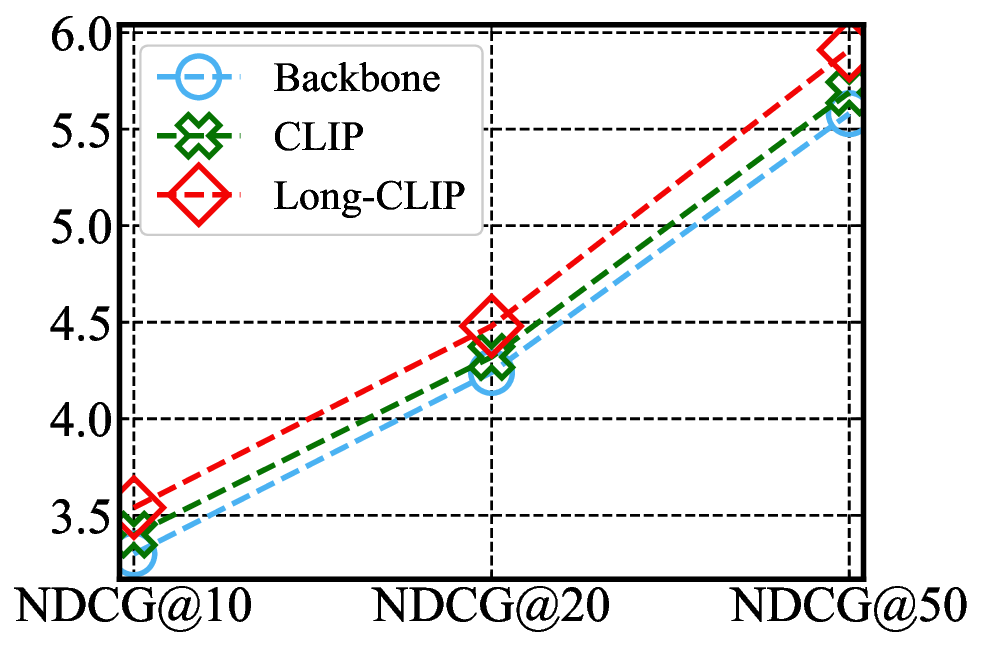}
	}
	\caption{Impact of CLIP.}
	\label{fig:clip choice}
\end{figure}

We employ the long-text favorable Long-CLIP \cite{zhang2024longclip} to further model the textual details.
To explore the effectiveness of different cross-modal alignment implementations, we conduct analysis studies on the Baby dataset in terms of clip implementations, as shown in Figure~\ref{fig:clip choice}.
Our observations are as follows:
Firstly, regardless of the CLIP version chosen, the CLIP-enhanced model consistently outperforms the vanilla backbone model, showcasing the efficacy of cross-modal semantic alignment.
Furthermore, compared to CLIP, Long-CLIP possesses an advantage in processing long texts (with a maximum length of $248$ tokens), enabling it to delve deeper into item details and further enhance recommendation performance.

\section{Parameter Sensitivity Study}
\label{sec: appendix temperature tau}
The similarity measurement scores attentively integrate each uni-view multi-modal representation for recommendation by incorporating the temperature coefficient $\tau$.
We search for the optimal value of $\tau$ in the range of $0.05, 0.1, 0.2, 0.5, 0.7$ and present the comparative analysis in Table \ref{table: temperature}.
In most scenarios, setting $\tau=0.1$ yields the best performance; however, there are certain cases where setting $\tau=0.05$ achieves superior results.
The performance of \cliper tends to deteriorate when $\tau > 0.1$.
We uniformly set $\tau = 0.1$ for similarity measurement.

\begin{table}[t]
	\caption{Impact of Temperature Coefficient $\tau$.}
	\resizebox{\linewidth}{!}{
		\begin{tabular}{c|ccc|ccc}
			\toprule
			$\tau$ & R@10          & R@20           & R@50           & N@10          & N@20          & N@50          \\
			\midrule
			0.05   & \textbf{6.94} & 10.28          & 16.09          & \textbf{3.73} & 4.58          & 5.74          \\
			0.1    & 6.89          & \textbf{10.37} & \textbf{16.24} & 3.69          & \textbf{4.58} & \textbf{5.75} \\
			0.2    & 6.89          & 10.19          & 16.01          & 3.67          & 4.51          & 5.67          \\
			0.5    & 6.83          & 10.18          & 16.00          & 3.67          & 4.52          & 5.68          \\
			0.7    & 6.89          & 10.16          & 15.99          & 3.68          & 4.51          & 5.68          \\
			\bottomrule
		\end{tabular}
	}
	\label{table: temperature}
\end{table}

\end{document}